\documentclass[twocolumn,aps,prl,showpacs,superscriptaddress]{revtex4}

\usepackage{graphicx}
\usepackage{amsmath,amssymb}
\usepackage{color}
\usepackage{epsfig}

\begin{document}


\title{Specker's fundamental principle of quantum mechanics}



\author{Ad\'an Cabello}
 \email{adan@us.es}
 \affiliation{Departamento de F\'{\i}sica Aplicada II, Universidad de
 Sevilla, E-41012 Sevilla, Spain}

\date{\today}


\date{\today}



\begin{abstract}
 I draw attention to the fact that three recently proposed physical principles, namely ``local orthogonality'', ``global exclusive disjunction'', and ``compatible orthogonality'' are not new principles, but different versions of a principle that Ernst Specker noticed long ago. I include a video of Specker stating this principle in 2009 in the following terms: ``Do you know what, according to me, is the fundamental theorem of quantum mechanics? (\ldots) That is, if you have several questions and you can answer any two of them, then you can also answer all of them''. I overview some results that suggest that Specker's principle may be of fundamental importance for explaining quantum contextuality. Specker passed away in December 10, 2011, at the age of 91.
\end{abstract}


\pacs{03.65.Ud,03.67.Mn,42.50.Xa}

\maketitle


\section{Introduction}


In a recent article entitled ``Quantum contextuality from a simple principle?'' \cite{Henson12}, Henson pointed out that in my paper ``A simple explanation of the quantum violation of a fundamental inequality'' \cite{Cabello12}, in which I presented an explanation of the maximum quantum violation of the Klyachko-Can-Binicio\u{g}lu-Shumovsky (KCBS) inequality \cite{KCBS08} (Result~1 in \cite{Cabello12}), and also pointed out that the proof of impossibility of Popescu-Rohrlich (PR) nonlocal boxes \cite{PR94} by Fritz {\em et al.}\ in \cite{FSABCLA12} can be connected to the fact that the graph representing the exclusivity structure of the Clauser-Horne-Shimony-Holt (CHSH) inequality \cite{CHSH69} contains, induced, exactly the same exclusivity structure of the KCBS inequality (Observation 1 in \cite{Cabello12}), and also introduced a family of quantum correlations exactly singled out by the principle (initially called ``global exclusive disjunction'') that the sum of probabilities of pairwise exclusive events cannot be higher than 1 (Result~2 in \cite{Cabello12}), I was implicitly assuming that ``if all pairs in a set of events are pairwise exclusive, [then] the set can itself be considered exclusive''. Indeed, I was making that assumption. Henson called this assumption ``consistent exclusivity''.

Later, Henson, in versions 2 and 3 of \cite{Henson12}, added a note pointing out that ``consistent exclusivity'' is, ``when the exclusive sets are defined with relevance to non-locality'', equivalent to the ``local orthogonality'' defined by Fritz {\em et al.}\ in \cite{FSABCLA12}. Henson also pointed out that the word `local' is inappropriate when applied to the KCBS scenario or general scenarios where different ``parties'' are not distinguished. Consequently, Henson proposed to call the principle ``compatible orthogonality''.


\section{Specker's principle}


The purpose of this note is to draw attention to the fact that this principle (in any of its formulations) is not new: Ernst Specker noticed it long ago and pointed out its fundamental importance. Moreover, this principle was already used to bound and single out quantum correlations. The story can be summarized as follows:
\begin{itemize}
 \item The origin can be traced back to Specker's 1960 paper ``The logic of propositions which are not simultaneously decidable'' \cite{Specker60} (cited as Ref. [9] in \cite{Cabello12}), where Specker shows how this assumption exactly singles out classical (and quantum; see below) correlations for a specific three-box game (later called Specker's parable of the overprotective seer \cite{LSW11}), while theories not satisfying this assumption can achieve higher values.
 \item It appears in the Kochen-Specker paper \cite{KS67}. Simon Kochen has recently commented the following: ``Ernst and I spent many hours discussing the principle (\ldots). The difficulty lays in trying to justify it on general physical grounds, without already assuming the Hilbert space formalism of quantum mechanics. We decided to incorporate the principle as an axiom in our definition of partial Boolean algebras. It appears on pp.~65--66 as follows: A partial Boolean algebra $C$ is a union of a family $F$ of Boolean algebras which is (i) closed under pairwise intersection of Boolean algebras, and such that (ii) if any two of a finite set $S$ of elements of $C$ lie in a common Boolean algebra in $F$ then all the elements of $S$ lie in a common Boolean algebra in $F$. (\ldots) I have never found a general physical justification for (ii)'' \cite{Kochen12}.
 \item Its importance as a fundamental principle was stressed by Specker in several occasions. For example, I heard it from Specker himself in June 17, 2009 in the following terms: ``Do you know what, according to me, is the fundamental theorem of quantum mechanics? (\ldots) That is, if you have several questions and you can answer any two of them, then you can also answer all three of them. This seems to me very fundamental''. It is recorded in video \cite{Specker09}. Two different fragments of that evening's recordings (one of them \cite{Specker09}) have been projected in several conferences since 2009. Longer fragments (including \cite{Specker09}) have been circulating among colleagues. The fragment \cite{Specker09} goes immediately after a comment that Specker makes on the three-box game of his 1960 paper, and before further details on what the principle means within the formalism of quantum mechanics.
 \item The importance of Specker's three-box game was also noticed in Liang, Spekkens, and Wisemans's paper significatively entitled ``Specker's parable of the overprotective seer: A road to contextuality, nonlocality and complementarity'' \cite{LSW11}, which was ``inspired by Ernst Specker's talk at the workshop `Information Primitives and Laws of Nature', which took place at ETH, Z\"urich in May 2008'' \cite{LSW11}, p.~32.
 \item In Ref.~\cite{CSW10}, Cabello, Severini, and Winter derived some consequences from assuming this principle, showing that it provides an upper bound and sometimes singles out quantum correlations. Although its fundamental role is not explicitly emphasized in \cite{CSW10}, this paper seems to be the inspiration of both Henson's \cite{Henson12} and Fritz {\em et al.}'s \cite{FSABCLA12}.
 \end{itemize}


\section{Boole, Specker, and Lov\'asz: An overview of some recent results}


In the following, I overview some results that suggest that the principle that pairwise decidable propositions are jointly decidable, together with Boole's condition that the sum of probabilities of jointly exclusive propositions cannot be higher than 1 \cite{Boole62}, which I will collectively call Specker's principle or simply S, may explain quantum contextuality. Specifically, by that here I mean explaining why the maximum quantum value for a convex combination of joint probabilities \cite{note1} of events whose exclusive disjunction structure is encoded in a graph $G$ in which events are represented by vertices and exclusive events are represented by adjacent vertices \cite{note2} is {\em exactly} given by the Lov\'asz number of $G$, $\vartheta(G)$, as shown in \cite{CSW10, CDLP12}. In \cite{CSW10}, it is shown that the maximum satisfying S is exactly given by the fractional packing number of $G$, $\alpha^* (G,\Gamma)$ when $\Gamma$ is the set of all cliques of $G$. In this case, we will denote it as $\alpha^* (G)$, while for general probabilistic theories $\Gamma$ is a general hypergraph. $\alpha^* (G,\Gamma)$ is defined as $\max \sum_{i\in V(G)} w_i$, where the maximum is taken over all $0 \leq w_i\leq 1$ and for all cliques $C_j \in \Gamma$, under the restriction $\sum_{i \in C_j} w_i \leq 1$. In order to single out structures with a quantum-classical separation, it is useful to recall that the maximum value for classical theories is given by the independence number of $G$, $\alpha(G)$ \cite{CSW10}. These tools allow us to summarize several old and recent results in a simple way.


\begin{figure}[t]
\begin{center}
\centerline{\includegraphics[scale=0.43]{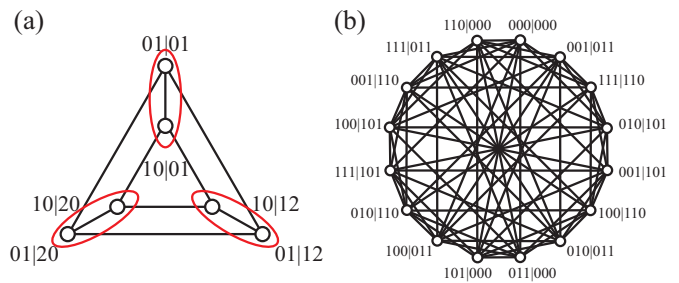}}
\caption{(a) Graph $G$ and (in red) hypergraph $\Gamma$ associated to Specker's three-box game.
(b) Graph $G$ associated to Mermin's inequality.}
\label{Fig1}
\end{center}
\end{figure}


{\em Specker's three-box game.---}What is the maximum value of the sum of probabilities that, if only two out of three boxes can be opened, one box is found empty and the other one is found full? If the three boxes are numbered 1, 2, and 3, and by $1,0|1,2$ we denote ``when boxes 1 and 2 are opened, box 1 is found full and box 2 is found empty'', this sum can be written as $\sum P(a,b|x,y)$, where $a,b\in \{0,1\}$, $a+b=1$, and $(x,y)=(1,2),(1,3),(2,3)$. The answer to the question for classical theories, quantum mechanics, and theories satisfying S is the same, since, using the language of \cite{CSW10}, $\alpha(C_3 \times K_2)=\vartheta(C_3 \times K_2)=\alpha^*(C_3 \times K_2)=2$, where $C_3 \times K_2$ is the 6-verte,x graph associated to the three-box game. This graph is shown in Fig. \ref{Fig1} (a). However, Specker noticed that in a world where S does not hold, this maximum can be $3$. In the language of \cite{CSW10}, this is so because $\alpha^*(C_3 \times K_2,\Gamma)=3$ for the $\Gamma$ resulting from the fact that the three boxes cannot be opened simultaneously and therefore only the propositions in the same red set in Fig. \ref{Fig1} (a) can be decided simultaneously.


{\em S provides an upper bound for quantum correlations.---}This immediately follows from the fact that, for any $G$, $\vartheta(G) \le \alpha^*(G)$ \cite{CSW10}.


{\em S singles out quantum correlations in all those cases in which $\vartheta(G)=\alpha^*(G)$.---}These include many important cases (see \cite{Cabello12} for a list). For instance, the operator of the Mermin inequality \cite{Mermin90} is equivalent to $\sum P(a,b,c|x,y,z)$, where the sum is extended to all $x,y,z\in \{0,1\}$ and $a,b,c\in \{0,1\}$ such that $a \oplus b \oplus c=xyz=0$, where $\oplus$ denotes sum modulo 2. $a,b,c|x,y,z$ is the event ``the results $a,b,c$ are respectively obtained when the pairwise separated tests $x,y,z$ are performed''. In this case, we have a quantum-classical separation, since $\alpha(CS)=3$ and $\vartheta(CS)=4$, where $CS$ denotes the complement of the Shrikhande graph and is shown in Fig. \ref{Fig1} (b). The maximum quantum value is singled out by S, since $\alpha^*(CS)=4$.


{\em S excludes PR boxes.---}This result was obtained in \cite{FSABCLA12}. In the language of \cite{CSW10}, this follows, e.g., from the fact that $\alpha^*[Ci_8(1,4) \ast Ci_8(1,4)] )=64 \frac{1}{5}$, where $\ast$ denotes or product, while PR boxes assign a higher value: $64 \frac{1}{4}$. $Ci_8(1,2)$ is the graph corresponding to $\sum P(a,b|x,y)$, where the sum is extended to all $x,y\in \{0,1\}$ and $a,b\in \{0,1\}$ such that $a \oplus b =xy$. On the other hand, the quantum value is $\vartheta[Ci_8(1,4) \ast Ci_8(1,4)]=64 \frac{6 + 4 \sqrt{2}}{64}$. Ref.~\cite{FSABCLA12} studies S in other multipartite scenarios.


{\em S might not be enough to single out $\vartheta(G)$.---}In \cite{FSABCLA12}, it is stated that ``Navascu\'es has a proof that the set of quantum correlations is strictly smaller than the set satisfying local orthogonality''. Without seing this proof, I cannot judge whether this affects the program stated before \cite{note1}.


{\em S singles out the quantum correlations for the KCBS inequality, which is the inequality associated to the simplest graph with quantum-classical separation.---}This is Result~1 in \cite{Cabello12}. In the language of \cite{CSW10}, it follows from the fact that $\vartheta(C_5 \ast C_5)=\alpha^*(C_5 \ast C_5)=25 \frac{1}{5}$.


{\em S singles out quantum correlations in all those cases in which $\vartheta(G \ast G)=\alpha^*(G \ast G)$.---}Result~2 in \cite{Cabello12} identifies graphs with this property.


{\em Every graph representing correlations with quantum-classical separation has, induced, (i) odd cycles of length 5 or more and/or (ii) their complements.---}This result was introduced in \cite{CDLP12b}. It is an open problem whether the quantum correlations of these families are singled out by S. However, the accumulated knowledge in graph theory is compatible with/suggests that the quantum correlations of family (i)/(ii) are singled out by S. In any case, for family (ii), the difference between correlations satisfying S and quantum correlations would be unappreciable in actual experiments. In the language of \cite{CSW10}, this follows from the fact that $\lim_{n \rightarrow \infty} \vartheta(\bar{C}_m^{\ast n}) \approx \lim_{n \rightarrow \infty} \alpha^*(\bar{C}_m^{\ast n})$, for $m$ odd $\ge 5$. If we could explain the maximum quantum contextuality of an arbitrary graph in terms of the maximum quantum contextuality of their induced elements of families (i) and (ii) \cite{CDLP12c}, then a proof that S singles out quantum correlations for families (i) and (ii) would be a strong evidence that S may be the fundamental principle of quantum contextuality.


Further investigations will solve this puzzle. The purpose of this note is to argue why, in my opinion, Specker deserves the credit for noticing this principle. It is also clear for me that the influence, direct or indirect, of his work is behind all these recent developments.


\begin{acknowledgments}
The author thanks Joe Henson for helpful discussions, Simon Kochen for very valuable feedback, and Daniel Schvartzman for help editing and subtitling \cite{Specker09}. This work was supported by the Project No.\ FIS2011-29400 (Spain).
\end{acknowledgments}




\begin{thebibliography}{99}
\bibitem{note1}
 Notice that I say ``a convex combination'' not ``any convex combination'', since my purpose here is to explain the maximum quantum correlations represented by $G$, not the maximum quantum value of any correlations represented by $G$. The following example illustrates the difference: the maximum quantum value for the $I_{3322}$ Bell inequality, represented by the graph $G(I_{3322})$, is smaller than the maximum quantum value that some correlations represented by the same graph $G(I_{3322})$ can attain. My purpose here is to understand this second maximum, not the first one. A method to obtain explicit quantum correlations reaching this second maximum is described in Ref.~\cite{CDLP12}.
\end{thebibliography}
\end{document}